\documentclass[11pt]{article}
\pdfoutput=1 

\usepackage[T1]{fontenc} 

\title{{\stb Entopy release in Electroweak Phase Transition in 2HDM  }}

\author{Arnab Chaudhuri\\Department of Physics and Astronomy, Novosibirsk State University,\\ email: arnabchaudhuri.7@gmail.com\\Maxim Yu. Khlopov\\Institute of Physics, Southern Federal University,\\ Université de Paris, CNRS, Astroparticule et Cosmologie and\\National Research Nuclear University "MEPHI" \\ email: khlopov@apc.in2p3.f\\Shiladitya Porey\\Department of Physics and Astronomy, Novosibirsk State University,\\ email: shiladityamailbox@gmail.com }

\usepackage{color,xcolor,graphicx,amsmath,amssymb,amsxtra,mathtools,hyperref,cleveref,subcaption,booktabs,multirow,geometry,physics, slashed, soul, ulem,cancel,tablefootnote}
\usepackage{paracol} 
\usepackage[export]{adjustbox}

\newcommand{\be}{\begin{eqnarray}}
\newcommand{\ee}{\end{eqnarray}}
\newcommand{\baln}{\begin{aligned}}
\newcommand{\ealn}{\end{aligned}}
\newcommand{\ban}{\begin{align}}
\newcommand{\ean}{\end{align}}
\newcommand{\summe}[2]{\sum\limits_{#1}^{#2}}

\newcommand{\unit}[1]{\mathrm{\ #1}}

\usepackage{auncial}
\usepackage[B1]{fontenc}

\newcommand{\lt}{\left }
\newcommand{\rt}{\right }

\definecolor{amber(sae/ece)}{rgb}{1.0, 0.49, 0.0}

\definecolor{blue(ncs)}{rgb}{0.0, 0.53, 0.74}

\usepackage{slashed}
\hypersetup{colorlinks,linkcolor={green!50!black},citecolor={cyan},urlcolor={red}}

\newcommand{\stb}{ \bf \color{blue!30!red} }

\definecolor{darkpastelgreen}{rgb}{0.01, 0.75, 0.24}
\def\l{\lambda}

\begin{document}
\maketitle

\begin{abstract}
Electroweak phase transition in the simplest extension of the standard model namely two Higgs doublet model and entropy production within this framework is studied. We have considered several benchmark points which were called using BSMPT, a C++ package, within the limit of $vev/T_C>0.2$ are studied, and corresponding entropy productions are shown in this paper.

\end{abstract}

\noindent Keywords: Electroweak symmetry breaking, Entropy.

\vspace{0.5cm}
\section{Introduction}
For a successful explanation about the origin of excess baryons over antibaryons in the universe
through electroweak baryogenesis (EWBG), 
a strong first-order electroweak phase transition (EWPT) in the early universe is necessary
. 
Cosmic EWPT happened when the hot universe cooled down enough in the primeval time so that the potential of the Higgs field got and settled at a non-zero minimum and in consequence, the symmetry of the theory $SU(2)_L\times U(1)_Y$ broke to $U(1)_{\rm em}$. At the time of first-order EWPT, bubbles of the broken phase originate and baryon-antibaryon asymmetry generates outside the wall of the bubbles of the broken phase.
However, after the discovery of the standard model (SM) Higgs boson, it is widely known that EWPT in SM with a single Higgs field is just a smooth cross-over. Therefore, for a successful EWBG, a theory of EWPT in 
beyond SM (BSM) is needed~\cite{Morrissey:2012db}.

On the other side, $\sim 26.5\%$ of the total energy density of the universe is contributed by the dark matter (DM) whose mysterious nature has not been unveiled till now. Although, primordial black holes and MACHOs which are considered as one of the viable baryonic DM candidates, it is now clear that they 
are unable to contribute completely to the 
DM energy density of the universe. There are theories about multicharged extension of the standard model like dark atoms which can be viable dark matter candidates, ~\cite{Chaudhuri:2021ppr}. But there are no experimental evidences as of now.

Not only about the baryogenesis, but there is also no irrefutable theory in SM about nonbaryonic DM particle which can successfully explain all the observations.

Similar to the above mentioned facts, there are many limitations of the SM. Thus scientists are desperately searching for experimental evidence of BSM. For them the recent result from Fermilab about $g_\mu -2$ for muon may be a ray of hope. $g_\mu$ is the gyromagnetic ratio of muon which is defined as the ratio of magnetic moment to the angular moment of muon and whose value is $2$ from tree-level calculation. If we define $a_\mu = (g_\mu-2)/2$, then higher order loop corrections from SM gives $a_\mu = 116,591,810(43)~\times~10^{-11}$ where the value measured from Fermilab is $16,592,061(41)~\times~10^{-11}$ which differs from SM at $3.3\sigma$ level~\cite{Muong-2:2021ojo}. This contradiction is actually buttressed the previously claimed result from the E821 experiment at Brookhaven National Lab (BNL). There are numerous explanation for this anomalous result including the existence of BSM.

Among all the BSM theories, the two Higgs doublet model (2HDM) is one of the most popular theories which not only exhibits strong first-order EWPT for the proper choice of parameter space but also 
provides the minimal phenomenological description of some effects of the supersymmetric model predicting two Higgs boson doublet.
 In addition to that, this model can produce dark matter particles~\cite{Wang:2014elb} and gives a satisfactory explanation for the $g_\mu$ anomaly of muon~\cite{Wang:2014sda} if the parameter space is properly chosen.

At or around the epoch of EWPT the energy density of the universe was dominated by relativistic species with negligible chemical potential. In addition to that, the universe was almost always in thermal equilibrium except some special epochs. Thus entropy density per comoving volume of the relativistic plasma was conserved. However, EWPT is a strongly thermally non-equilibrium process and thus there is a possibility that entropy might have been generated during this cosmic process.


In this work, we explored the increase in entropy during the epoch of EWPT in the real type-I 2HDM framework. We have shown that entropy density per comoving volume increases if EWPT happens as a first-order phase transition in 2HDM model. 

The article is arranged as follows: In the next section the Lagriangian of the model along with the results are given. A generic conclusion follows and in the appendix the detailed potential is mentioned.

\section{Lagrangian density of the model}
The Lagrangian density of EWPT theory in real type-I 2HDM is given by
\be
\mathcal{L}=\mathcal{L}_{\rm gauge, kin} + \mathcal{L}_f+ \mathcal{L}_{\rm Yuk}  + \mathcal{L}_{\rm Higgs} - V
(\Phi_1, \Phi_2,  T)
\ee 
where $\mathcal{L}_{\rm gauge, kin}$, $\mathcal{L}_f$ and $\mathcal{L}_{\rm Yuk}$ are the kinetic energy term of gauge bosons ($W_\alpha$ and $B_\alpha$ with $\alpha=0,1,2,3$) , kinetic energy of fermions and Yukawa interaction term of fermions with Higgs bosons.  These terms are defined in~\cite{Chaudhuri:2021agl,Chaudhuri:2021rwt} and also discussed in Appendix~\ref{appendixA}. Throughout this article, all the Greek indices used in super or sub-script run from $0$ to $3$ and Latin indices from $1$ to $3$ if not mentioned otherwise.

$\mathcal{L}_{\rm Higgs}$ incorporates the kinetic term of the Higgs field and their interaction with the gauge bosons. Thus
\be
\mathcal{L}_{\rm Higgs}= \lt\{(\partial^\mu  + i {\cal W}^\mu 
) \Phi_a \rt\}^{\dagger} \lt\{(\partial_\mu  + i {\cal W}_\mu 
) \Phi_a \rt\}
\ee 
where $a=1,2$ for two Higgs field, $i=\sqrt{(-1)}$, $i {\cal W}^\mu \equiv + ig T^k {W^k}^\mu + i g'Y B^\mu$, and 
 $g$ and $g'$ are coupling constants, $T^i$ is the generator of $SU(2)_L$ (left-Chiral), which is also a form of Pauli matrices, and $Y$ is the hyper-charge generator of the $U(1)$.

The total CP-conserving potential for our 2HDM model considered is
\be
 V(\Phi_1,\Phi_2,T)=V_{tree}(\Phi_1,\Phi_2)+V_{CW}(\Phi_1,\Phi_2)+V_{T}(T) {+ V_{\rm daisy} (T)}    \label{eq:Total potential}
\ee 
The tree-level potential can be written as 
\begingroup\makeatletter\def\f@size{10}\check@mathfonts
\def\maketag@@@#1{\hbox{\m@th\normalsize\normalfont#1}}%
\begin{align} 
V_{\rm tree}(\Phi_1,\Phi_2)   =& m_{11}^2 \Phi_1^\dagger \Phi_1 + m_{22}^2 \Phi_2^\dagger \Phi_2 - \left[m_{12}^2 \Phi_1^\dagger \Phi_2 + m_{12}^* \Phi_2^\dagger \Phi_1 \right] + \frac{1}{2} \lambda_1 \left(\Phi_1^\dagger \Phi_1\right)^2 \nonumber \\ 
&+  \frac{1}{2} \lambda_2 \left(\Phi_2^\dagger \Phi_2\right)^2  + \lambda_3 \left(\Phi_1^\dagger \Phi_1\right)\left(\Phi_2^\dagger \Phi_2\right) + \lambda_4 \left(\Phi_1^\dagger \Phi_2\right)\left(\Phi_2^\dagger \Phi_1\right) \\
&+ \left[\frac{1}{2} \lambda_5  \left(\Phi_1^\dagger \Phi_2\right)^2 + \frac{1}{2} \lambda_5^*  \left(\Phi_2^\dagger \Phi_1\right)^2 \right]\nonumber .
\end{align}
\endgroup

$m_{12}^2$, $m_{11}^2$, and $m_{22}^2$ can be estimated from the following formula 
\be
m_{12}^2 &=& 100^2 \unit{GeV}^2 \\
m_{11}^2 &=& \frac{1}{4 v_1} \lt(-2 \lambda_1 v_1^3 + 4 m_{12}^2 v_2 - 2 \lambda_3 v_1 v_{2}^2 - 2 \lambda_4 v_1 v_2^2 - \lambda_5 v_1 v_2^2 -  v_1 v_2^2 \lambda_5 \rt) \\
m_{22}^2 &=& \frac{1}{4 v_2} \lt( 4 m_{12}^2 v_1 - 2 \lambda_3 v_1^2 v|_2 - 2 \lambda_4 v_1^2 v_2 - \lambda_5 v_1^2 v_2 - 2 \lambda_2 v_2^3 - v_1^2 v_2 \lambda_5 \rt)
\ee 
The value of $m_{12}^2$ can alter for different parameter space. These formulas are valid since $\lambda_5$ is real and $\lambda_6=\lambda_7=0$. The $\lambda_{1-5}$ can be calculated from the parameter space as
\be
\l_1 &=& \frac{m_{H}^2 \cos{\alpha}^2 + m_h^2 \sin\alpha^2 - m_{12}^2 \tan\beta}{v^2 \cos\beta^2} , \\
\l_2 &=& \frac{ m_{H}^2 \sin\alpha^2 + m_h^2 \cos{\alpha}^2 - m_{12}^2 \tan\beta^{-1} }{v^2 \sin\beta^2}  , \\
\l_3 &=& \frac{ (m_{H}^2 - m_h^2) \sin\alpha \cos{\alpha} + 2 m_{H^{\pm}}^2 \cos\beta \sin\beta - m_{12}^2}{v^2 \sin\beta \cos\beta} , \\
\l_4 &=& \frac{ (m_A^2 - 2 m_{H^{\pm}}^2) \sin\beta \cos\beta + m_{12}^2}{v^2 \sin\beta \cos\beta} , \\
\l_5 &=& \frac{ m_{12}^2 - m_A^2 \sin\beta \cos\beta}{v^2 \sin\beta \cos\beta}  .
\label{sec2:masstophys}
\ee 
where $v$ is the standard model expectation value, $v^2 = v_1^2 + v_2^2$, $\tan\beta=v_2/v_1$ and $\cos(\beta-\alpha)\to 0$ leads to SM result. The details about the parameter space of $m_H$, $m_h$, $m_{H\pm}$ can be found in the recent works~\cite{Eberhardt:2020dat,Karmakar:2020mds,Dorsch:2016tab}.

The Coleman–Weinberg correction to the potential -
\be 
V_{\rm CW}\left(v_1+v_2 \right)=\sum_j \frac{n_j}{64\pi^2} (-1)^{2s_j}m_j^4\left(v_1,v_2\right)\left[ \log\left( \frac{m_j^2 \left(v_1,v_2 \right)}{\mu^2} \right) - c_j \right] 
\ee 

The values of $n_j$, $s_j$, $c_j$ and different mass-values $m_j^2\left(v_1,v_2 \right)$ are mention in Appendix~\ref{s-a} and $\mu = 246 \unit{GeV}$.

Temperature correction of potential and its series expansion in Landau gauge are 

\be 
&& V_T =\frac{T^4}{2\pi^2}\left( \sum_{j={\rm bosons}} n_j J_{B}\left[\frac{m_j^2(v_1,v_2)}{T^2}\right]  + \sum_{j={\rm fermions}} n_j J_{F}\left[\frac{m_j^2(v_1,v_2)}{T^2}    \right] \right) \\
&&T^4 J_B \left[\frac{m^2}{T} \right] = -\frac{\pi^4 T^4}{45} + \frac{\pi^2}{12}T^2 m^2 - \frac{\pi}{6}T (m^2)^{3/2} - \frac{1}{32}m^4 \ln \frac{m^2}{a_b T^2} + \cdots,   \\
&&T^4 J_F \left[\frac{m^2}{T} \right] =  \frac{7\pi^4 T^4}{360} - \frac{\pi^2}{24}T^2 m^2 - \frac{1}{32}m^4 \ln \frac{m^2}{a_f T^2} + \cdots, 
\ee 
 where $a_b=16a_f=16\pi^2 \exp(3/2-2\gamma_E)$ with $\gamma_E$ being the Euler-Mascheroni constant.

The daisy term is defined as 
\be
V_{\text{daisy}}(T) &= -\frac{T}{12\pi} \summe{i=1}{}
\left[ 
\lt(M_i^2\lt(v_1, v_2, T \rt)\rt)^{3/2} - \lt(m_i^2\lt(v_1, v_2 \rt)\rt)^{3/2}%
\right]  \label{eq:AE2} 
\ee
Details about the $M_i^2\lt(v_1, v_2, T \rt)$ term can be found in~\cite{Bernon:2017jgv, Basler:2018cwe}. Actually, we will see later that all these terms will be taken care of the software package we have used for this work.

At sufficiently high temperature, the total potential of eq.\eqref{eq:Total potential} has only one minimum at $\lt<\Phi_1\rt>=\lt<\Phi_2\rt>=0$ and there is no symmetry breaking. The critical temperature ($T_c$) is defined as the temperature at which if the temperature drops down, the total potential gets a second minimum at $(\Phi_{a,{\rm min}}) \equiv \lt\{ \lt<\Phi_1\rt>=v_1, \lt<\Phi_2\rt>=v_2 \rt\}$. For simplicity, we are assuming in this work that both of the Higgs field $\Phi_1$ and $\Phi_2$ get the second minimum
 at the same temperature $T_c$ at the same time. Thus
\be
V\left(\Phi_1=0,\Phi_2=0,T_c \right)= V\left(\Phi_1=v_1,\Phi_2=v_2,T_c \right).
\ee 

As soon as the Higgs potential gets a non-zero minimum, the other relativistic particles starts to gain mass and becomes non-relativistic. The reaction rate among them and also with photon becomes comparable with the Hubble parameter and thus decouples from relativistic plasma. The mass of the particle and coupling constant determine the decoupling temperature. For instance, top quark decouples earlier than electron or other quarks. 

Now, at the time of EWPT 
the universe can be assumed as perfectly homogeneous and isotropic and thus we can neglect the spatial partial derivatives of the Higgs fields.  
Therefore, when the Higgs fields start to oscillate around their minima $(\Phi_{a,{\rm min}})$ then energy density $\rho$ and pressure $P$ are
\be
\rho &=& \Dot{\Phi}_{a, {\rm min}}^2
+  V_{\rm tot}(\Phi_1,\Phi_2, T) + \frac{g_* \pi^2}{30} T^4.  \label{eq:energy density}  \\
P &=& \Dot{\Phi}_{a, {\rm min}}^2 -  V_{\rm tot}(\Phi_1,\Phi_2, T) + \frac{1}{3} \frac{g_* \pi^2}{30} T^4  \label{eq:pressure density} 
\ee 
 The 
 last terms in eq.\eqref{eq:energy density} and eq.\eqref{eq:pressure density}  arise from the Yukawa interaction between fermions and Higgs bosons and from
the energy density of the fermions, the gauge bosons, and the interaction between the Higgs and gauge bosons.
 $g_*$ depends on the effective number of particles present in the relativistic soup at or near the EWPT. It's value in our model is greater than the value in SM.


Since the oscillation of $\Phi_a$ around $\Phi_{a, {\rm min}}$ is small compared to Hubble expansion, we can neglect the time derivative of $\dot{\Phi}_{a,{\rm min}}$~\cite{Chaudhuri:2017icn} for simplicity in this work.

\begin{figure}[htp]
  \centering
  \begin{minipage}[b]{0.4\textwidth}
    \fbox{\includegraphics[width=\textwidth]{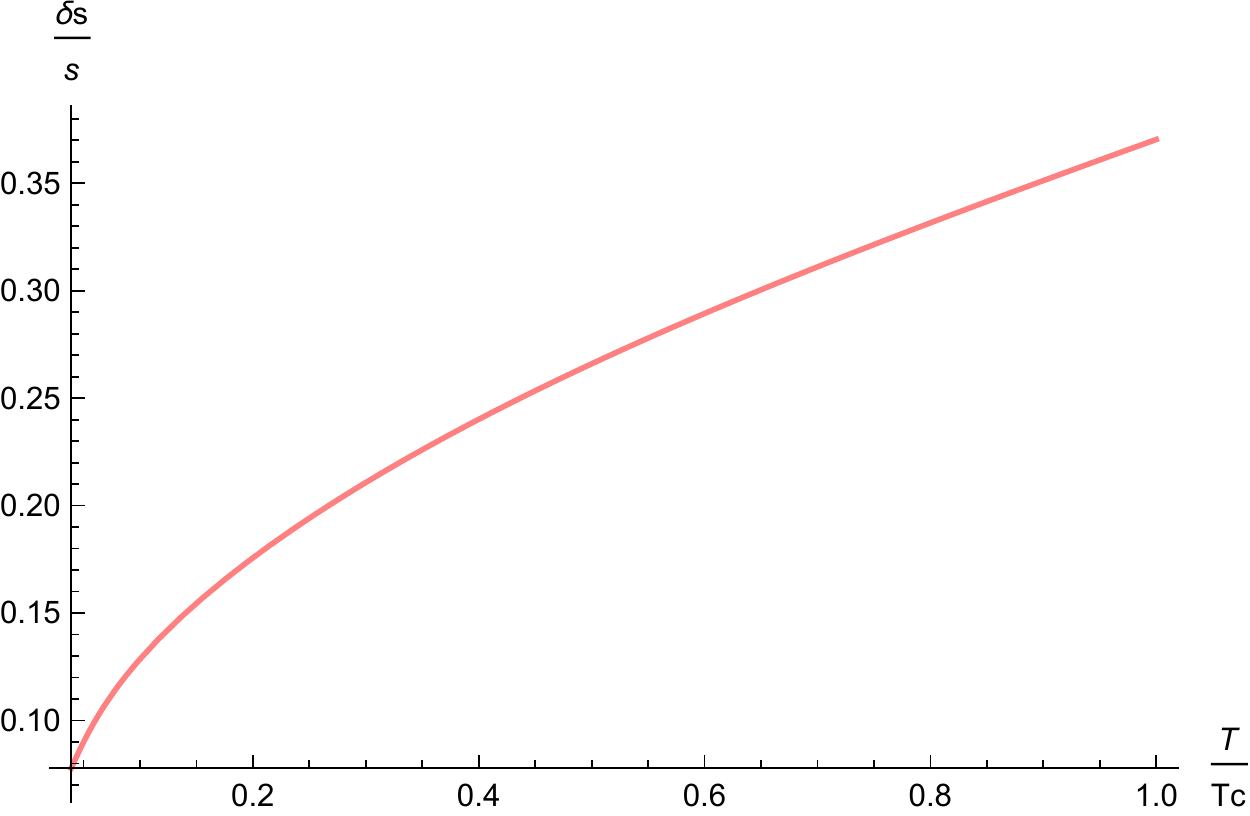}}
  \end{minipage}
  \hspace*{.1cm}
  \begin{minipage}[b]{0.4\textwidth}
    \fbox{\includegraphics[width=\textwidth]{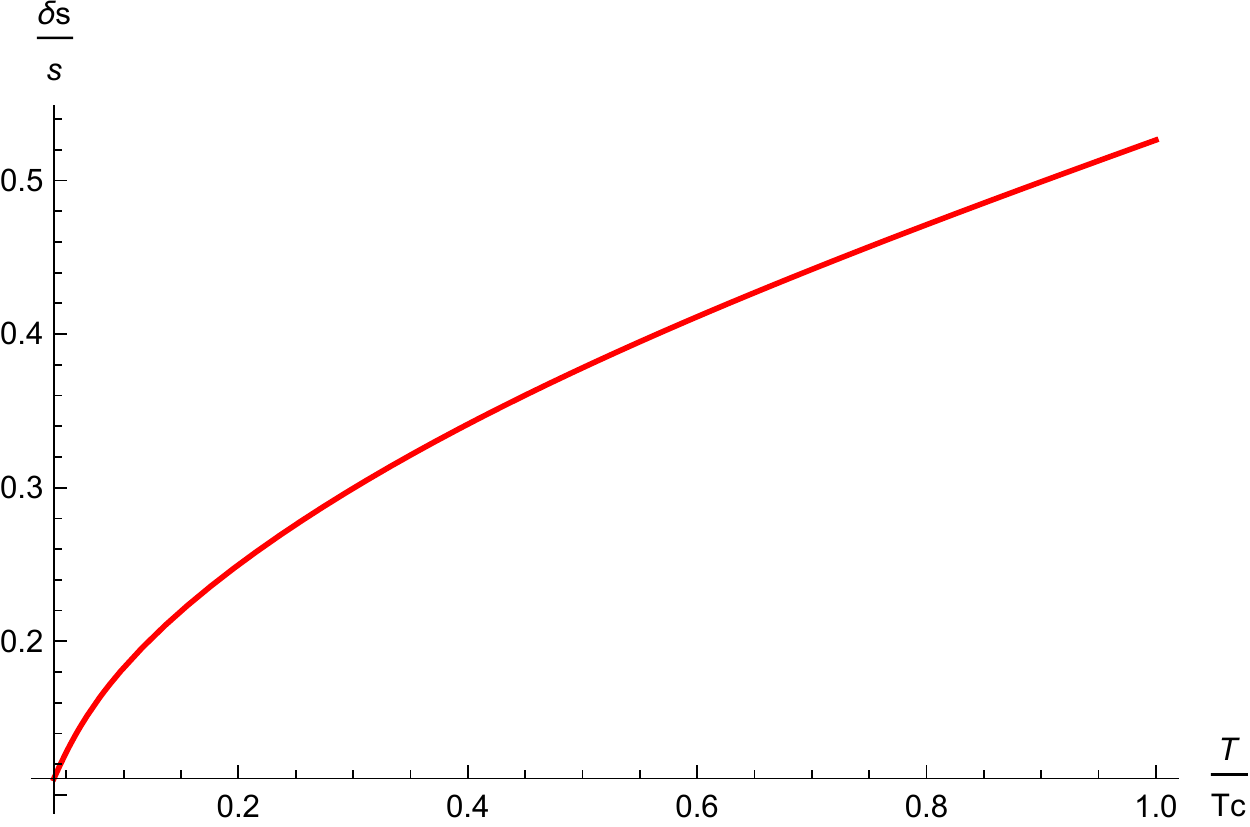}}
      \end{minipage}
    \hspace*{.1cm}
    \begin{minipage}[b]{0.4\textwidth}
    \fbox{\includegraphics[width=\textwidth]{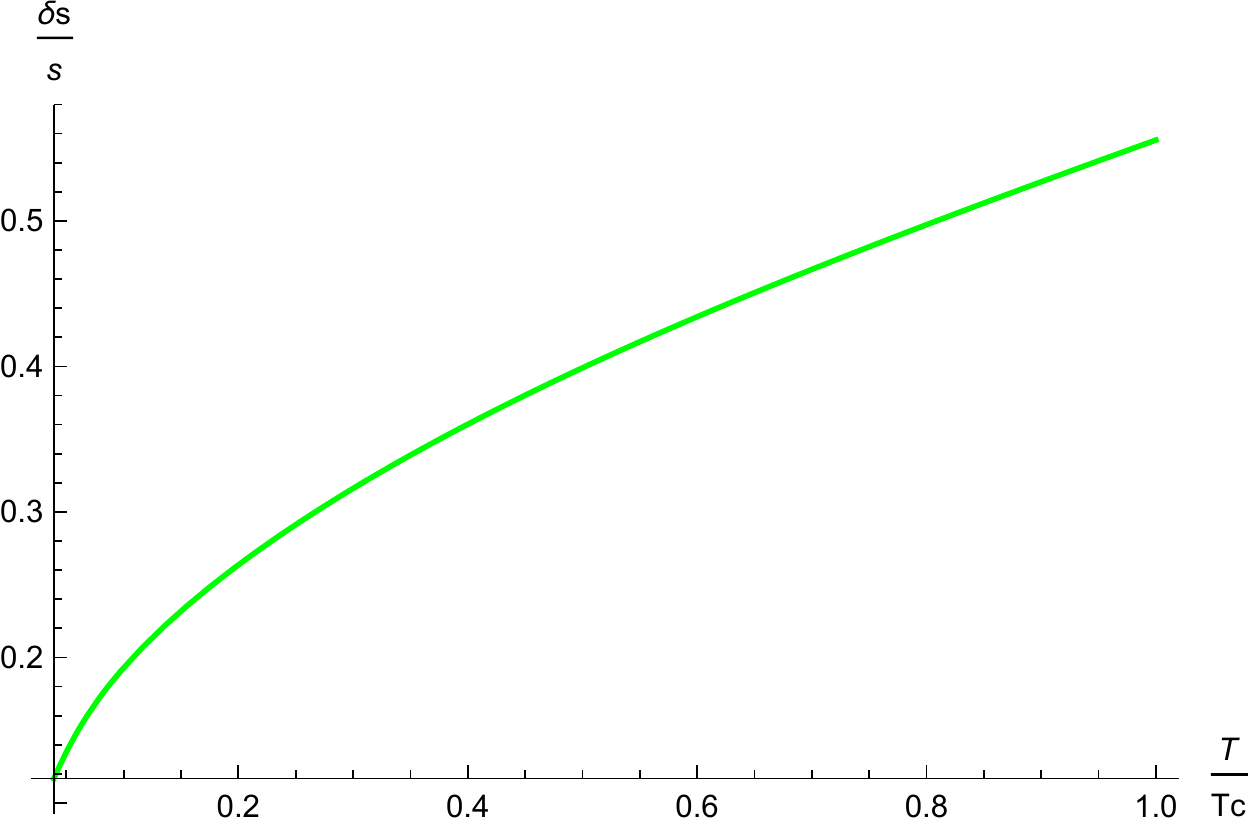}}
      \end{minipage}
      \begin{minipage}[b]{0.4\textwidth}
    \fbox{\includegraphics[width=\textwidth]{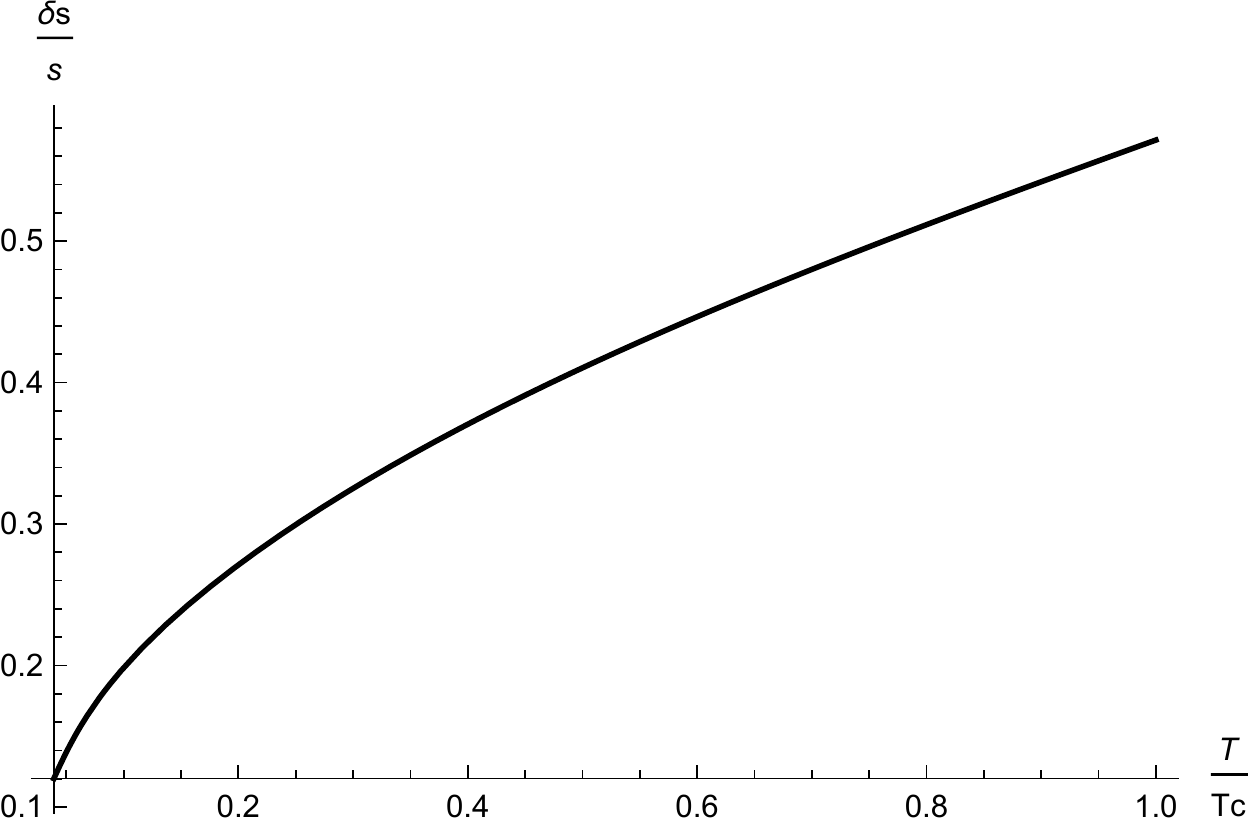}}
      \end{minipage}
      \begin{minipage}[b]{0.4\textwidth}
    \fbox{\includegraphics[width=\textwidth]{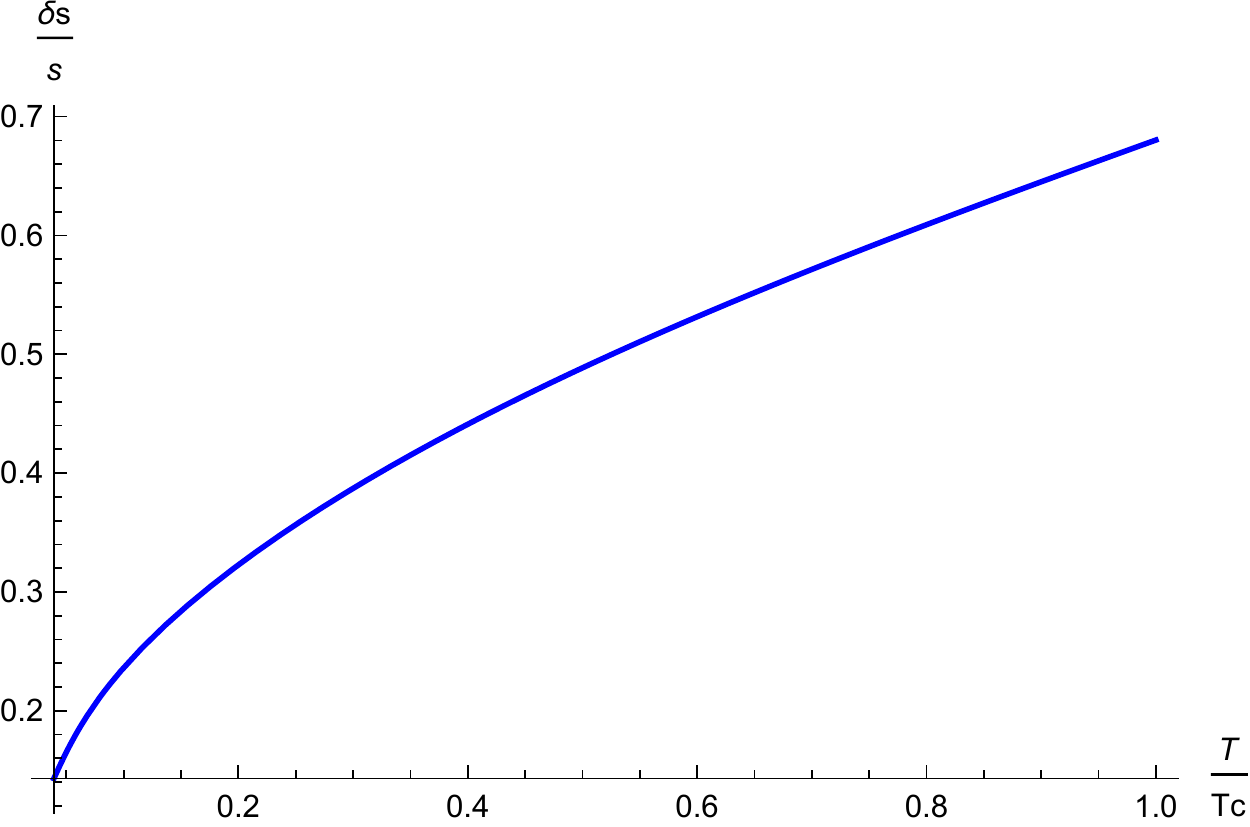}}
      \end{minipage}
  \caption{The figures show the entropy production for five different benchmark (BM) points: Pink line (BM 5, $T_c=135.38~\text{GeV}$ and $\delta s/s=37\%$), Red line (BM 2,  $T_c=153.27~\text{GeV}$ and $\delta s/s=53\%$), Green line (BM 1,  $T_c=161.36~\text{GeV}$ and $\delta s/s=30\%$), Black line (BM 3,  $T_c=168.61~\text{GeV}$ and $\delta s/s=59\%$), and Blue line (BM 4  $T_c=230.18~\text{GeV}$ and $\delta s/s=70\%$).}
 \label{f-entropy-1}
 \end{figure}

Again, entropy density per comoving volume is defined as 
\be
s= \frac{\rho + P}{T} a^3
\ee 
which is conserved for relativistic species with negligible chemical potential. 
From eq.\eqref{eq:energy density} and eq.\eqref{eq:pressure density} we get
\be
\rho + P = 2 \Dot{\Phi}_{a, {\rm min}}^2  + \frac{4}{3} \frac{g_* \pi^2}{30} T^4 
\ee 
As discussed earlier, $g_*$ will change with the decoupling process and thus $s$ for relativistic plasma will increase for our considered scenario. Then the increase in entropy can be calculated using conservation of energy momentum tensor
\be
\dot{\rho}= - 3 H \lt( \rho + P \rt)  \label{eq: solve it}
\ee 
To solve eq.\eqref{eq: solve it}, we have used BSMPT~\cite{Basler:2018cwe, Basler:2020nrq}, a C++ package to calculate the vacuum expectation value (VEV) of the total potential, value of the total potential at VEV for different temperatures including $T_c$. We have chosen the parameter in such a way so that $VEV/T_c>0.02$.  We have considered five sets of benchmark values and the corresponding figures are shown in Fig.\ref{f-entropy-1}


\begin{table}[h!] 
	\centering
	\begin{center}
		\caption{2HDM Benchmark points for entropy production}\label{Table:1 Benchmark values}
		\resizebox{0.99\textwidth}{!}{\begin{tabular}{ |c|c|c|c|c|c|c|c|c|c|c|c|c|c|c|c|} 
			\hline
			& \boldmath{$m_h$} \textbf{[GeV]}& \textbf{\boldmath{$m_H$} [GeV]} & \textbf{\boldmath{$m_{H^{\pm}}$} [GeV]}& \textbf{\boldmath{$m_A$} [GeV]} & \boldmath{$\tan\beta$} & \boldmath{$\cos \left(\beta-\alpha \right)$} & \textbf{\boldmath{$m_{12}^2~\text{GeV}^2$}} & \boldmath{$\lambda_1$} & \boldmath{$\lambda_2$} & \boldmath{$\lambda_3$} & \boldmath{$\lambda_4$} & \boldmath{$\lambda_5$} & \boldmath{$T_c$} & \boldmath{$vev/T_c$} & \boldmath{$\delta s/s[\%]$}\\
			\midrule
BM 1 & $125$ & $500$ & $500$ & $500$ & $2$ & $0$ & $10^5$ & $0.258$ & $0.258$ & $0.258$ & $0$ & $0$ & $ 161.36$ & 1.4 & $57$ \\
	\midrule
	\midrule
BM 2 & " & " & $485$ & $500$ & $2$ & $0.00$ & $10^5$ & $0.258$ & $.258$ & $-0.23$ & $0.49$ & $0$ & $153.27 $ & 1.25 & $53$ \\
	\midrule
	\midrule
BM 3 & " & " & $485$ & $485$ & $2$ & $0.07$& $10^5$ & $1.28$ & $0.002$ & $0.21$ & $0.244$ & $0.244$  & ${ 168.61}$ & 1.7 & $59$ \\
	\midrule
 BM 4 & " &  $485$ & $485$ & $485$ & $10$ & $0.1$& 23,289.6 & $3.9$ & $0.22$ & $3.9$ & $0$ & $0$  & ${ 230.18 }$  & 1.86 & $70$ \\
	\midrule
  BM 5 & " &  $90$ & $200$ & $300$ & $10$ & $0$ & $801.98$ & $0.258$ & $0.258$ & $1.31$ & $0.3$ & $-1.35$ & ${ 135.38  }$ & 1.06 & $37$  \\
	\midrule
		\end{tabular}}
	\end{center}
\end{table}


\section{Conclusion}
As seen from Fig.\ref{f-entropy-1}, the entropy productions for some benchmark points are shown here. A proper difference can be noticed from the standard model scenario. As seen in \cite{Chaudhuri:2017icn}, the entropy released is around $~13 \%$ and in the present scenario, we see that the production is considerably higher. This is because first-order phase transition as seen in 2HDM can release more entropy compared to smooth crossover or second-order in the case of the standard model. The massive scalar particles in 2HDM contribute considerably to this production as well.

\section*{Acknowledgements}
The work of S.P. and A.C. is funded by RSF Grant 19-42-02004. The research by M.K. was supported by the Ministry of Science and Higher Education of the Russian Federation under Project "Fundamental problems of cosmic rays and dark matter", No. 0723-2020-0040.

 {}

\section*{Appendix A}
\makeatletter\def\@currentlabel{A}\makeatother
\label{appendixA}

The kinetic energy term of gauge bosons, kinetic energy of fermions and Yukawa interaction term of fermions with Higgs bosons are 
\begin{align} 
&\mathcal{L}_f =\sum_{\Psi=Q,L,u,d,l}i \lt(\bar{\Psi}_L \slashed{D}\Psi_L + \bar{\Psi}_R \slashed{D}\Psi_R \rt) \\ 
&\mathcal{L}_{\rm Yuk} = -\left[ y_e \bar{e_R} \Phi_a^\dagger L_L + y_e^* \bar{L_L} \Phi_a^\dagger e_R  + \cdots \right]  \label{eq:Yukawa lagrangian}\\
%
%
&\mathcal{L}_{\rm gauge, kin} = -\frac{1}{4}G^j_{\mu \nu}{G^j}^{\mu \nu}-\frac{1}{4}F^B_{\mu \nu}{F^B}^{\mu \nu}  \label{eq:gauge kin}
\end{align}
where $\Psi$ is the fermionic field, subscript $L$ ($R$) is for the left (right) chiral field. The sum in eq.\eqref{eq:Yukawa lagrangian} is also over quarks. $y_e$ is the complex constant and
\begin{align}
    & G^j_{\mu \nu}=\partial_\mu W^j_\nu-\partial_\nu W^j_\mu - g \epsilon^{jkl}W_\mu^k W_\nu^l     \label{eq:Gmunu}\\
    & F^B_{\mu \nu}=\partial_\mu B_\nu-\partial_\nu B_\mu \\
    & \slashed{D} \Psi^{(j)}_{L,R} \equiv \gamma^\mu (\partial_\mu + ig W_\mu + i g'Y_{L,R} B_\mu) \Psi^{(j)}_{L,R}
\end{align}

\section{Appendix B: Masses of new Scalars}
\makeatletter\def\@currentlabel{B}\makeatother
\label{s-a}
\be
c_i=
\begin{cases}
      \frac{5}{6}, & \left(i=W^\pm, Z, \gamma \right)\\
      \frac{3}{2}, & \text{otherwise}
  \end{cases}
\ee

\begin{center}
\begin{tabular}{ | c | c| c | c| c| } 
\hline
Bosons & $n_i$ & $s_i$ & $m(v)^2$ & \\
\hline
$h$ & $1$ & $1$ & eigenvalues of \ref{Mass_of_neutral_Higgs_bosons} &  Higgs \\
\hline
$H$ & $1$ & $1$ & eigenvalues of \ref{Mass_of_neutral_Higgs_bosons} & Higgs \\
\hline
$A$& $1$ & $1$ & eigenvalues of \ref{Mass_of_neutral_Higgs_bosons} & Higgs\\
\hline
$G^0$& $1$ & $1$ & eigenvalues of \ref{Mass_of_neutral_Higgs_bosons} & Goldstone\\
\hline
$H^\pm$& $2$ & $1$ & Eq.\ref{Mass_of_mHPlusMinus}& Charged  Higgs\\
\hline
$G^\pm$& $2$ & $1$ & Eq.\ref{Mass_of_mGPlusMinus}& Charged Goldstone\\
\hline
$Z_L$& $1$ & $1$ & Eq.\ref{Mass_of_Z} & Higgs\\
\hline
$Z_T$& $2$ & $2$ & Eq.\ref{Mass_of_Z} & Higgs\\
\hline
$W_L$& $2$ & $1$ & Eq.\ref{Mass_of_W} & Higgs\\
\hline
$W_T$& $4$ & $2$ & Eq.\ref{Mass_of_W} & Higgs\\
\hline
$\gamma_L$& $1$ & $2$ & Eq.\ref{Mass_of_gamma} & \\
\hline
$\gamma_T$& $2$ & $2$ & Eq.\ref{Mass_of_gamma} & \\
\hline
\end{tabular}
\end{center}

\be
&& m_W^2=\frac{g^2}{4}v^2. \label{Mass_of_W}\\
&& m_Z^2=\frac{g^2+g'^2}{4}v^2. \label{Mass_of_Z} \\
&& m_\gamma^2=0. \label{Mass_of_gamma}
\ee

\be
\Bar{m}_{H^\pm}^2 &&=\frac{1}{2} \left( \mathcal{M}_{11}^C + \mathcal{M}_{22}^C \right)  +\frac{1}{2} \sqrt{4\left( \left( \mathcal{M}_{12}^C\right)^2 + \left( \mathcal{M}_{13}^C \right)^2 \right)+\left(\mathcal{M}_{11}^C-\mathcal{M}_{22}^C \right)^2}.  \label{Mass_of_mHPlusMinus} \\
\Bar{m}_{G^\pm}^2 &&=\frac{1}{2}\left( \mathcal{M}_{11}^C + \mathcal{M}_{22}^C +  \right) -\frac{1}{2} \sqrt{4\left( \left( \mathcal{M}_{12}^C\right)^2 + \left( \mathcal{M}_{13}^C \right)^2\right) +\left(\mathcal{M}_{11}^C-\mathcal{M}_{22}^C  \right)^2 }.    \label{Mass_of_mGPlusMinus}
\ee
where
\be
c_1 &&=\frac{1}{48}\left(12\lambda_1 + 8 \lambda_3 + 4 \lambda_4 + 3 \left( 3g^2 + g'^2\right)\right) \\
c_2 &&= \frac{1}{48}\left(12\lambda_2 + 8 \lambda_3 + 4 \lambda_4 + 3 \left( 3g^2 + g'^2\right) + \frac{24}{v_2^2}m_t^2(T=0)\right) \nonumber \\
&&+\frac{1}{2v_2^2}m_b^2(T=0)
\ee
where $m_t(T=0)=172.5 {\rm Gev}$ and $m_{b}(T=0)=4.92 {\rm GeV}$.
For our case $(v_3=0)$,
\be
\mathcal{M}_{11}^C &&=m_{11}^2 +\lambda_1 \frac{v_1^2}{2} + \lambda_3 \frac{v_2^2}{2}  \\
\mathcal{M}_{22}^C &&=m_{22}^2 +\lambda_2 \frac{v_2^2}{2} + \lambda_3 \frac{v_1^2}{2}  \\
\mathcal{M}_{12}^C &&=\frac{v_1v_2}{2}\left(\lambda_4+\lambda_5\right)-m_{12}^2 \\
\mathcal{M}_{13}^C &&=0
\ee

Masses of $h$, $H$ and $A$ are the eigen values of the matrix
\be
\Bar{\mathcal{M}}^N=\left( \mathcal{M}^N \right) \label{Mass_of_neutral_Higgs_bosons} 
\ee
For our case $(v_3=0)$,
\be
\mathcal{M}_{11}^N &&= m_{11}^2+\frac{3\lambda_1}{2}v_1^2+ \frac{\lambda_3 +\lambda_4}{2} v_2^2 + \frac{1}{2} \lambda_5 v_2^2\\
\mathcal{M}_{22}^N &&= m_{11}^2+\frac{\lambda_1}{2}v_1^2+ \frac{\lambda_3 +\lambda_4}{2} v_2^2 - \frac{1}{2} \lambda_5 v_2^2\\
\mathcal{M}_{33}^N &&= m_{22}^2+\frac{3\lambda_2}{2}v_2^2 + \frac{1}{2}\left(\lambda_3+\lambda_4 +\lambda_5 \right)v_1^2\\
\mathcal{M}_{44}^N &&= m_{22}^2+\frac{\lambda_2}{2}v_2^2 + \frac{1}{2}\left(\lambda_3+\lambda_4 -\lambda_5 \right)v_1^2\\
\mathcal{M}_{12}^N &&= 0\\
\mathcal{M}_{13}^N &&=-m_{12}^2+\left(\lambda_3+\lambda_4+\lambda_5\right) v_1 v_2 \\
\mathcal{M}_{14}^N &&=0 \\
\mathcal{M}_{23}^N &&=0 \\
\mathcal{M}_{24}^N &&= -m_{12}^2+\lambda_5 v_1 v_2\\
\mathcal{M}_{34}^N &&= 0
\ee

\begin{table}
\caption{Field dependent mass of all fermions}\label{Table: mass of all fermion}
\begin{center}
\begin{tabular}{ | c | c| c | c| c| }  
\hline
Fermions & $n_i$ & $s_i$ & $m_f(T=0)$ & \\
\hline
$e$ & $4$ & $\frac{1}{2}$ & $\frac{y_e}{\sqrt{2}}v_k$ & lepton\\

$\mu$ & $4$  & $\frac{1}{2}$ & $\frac{y_\mu}{\sqrt{2}}v_k$ & lepton\\

$\tau$ & $4$ & $\frac{1}{2}$ & $\frac{y_\tau}{\sqrt{2}}v_k$ & lepton\\
\hline
$u$ & $12$ & $\frac{1}{2}$ & $\frac{y_u}{\sqrt{2}}v_k$ & quark\\
$c$ & $12$ & $\frac{1}{2}$ & $\frac{y_c}{\sqrt{2}}v_k$ & quark\\
$t$ & $12$  & $\frac{1}{2}$ & $\frac{y_t}{\sqrt{2}}v_k$ & quark\\
$d$ & $12$ & $\frac{1}{2}$ & $\frac{y_d}{\sqrt{2}}v_k$ & quark\\
$s$ & $12$ & $\frac{1}{2}$ & $\frac{y_s}{\sqrt{2}}v_k$ & quark\\
$b$ & $12$ & $\frac{1}{2}$ & $\frac{y_b}{\sqrt{2}}v_k$ & quark\\
\hline
\end{tabular}
\end{center}
\end{table}


\begin{thebibliography}{99}
 
\bibitem{Morrissey:2012db}
D.~E.~Morrissey and M.~J.~Ramsey-Musolf,
New J. Phys. \textbf{14} (2012), 125003
doi:10.1088/1367-2630/14/12/125003
[arXiv:1206.2942 [hep-ph]].

\bibitem{Chaudhuri:2021ppr}
A.~Chaudhuri and M.~Yu.~Khlopov,
Universe \textbf{7}, (2021) 8, 275
doi:10.3390/universe7080275
[arXiv: 2106.11646 [hep-ph]].


\bibitem{Muong-2:2021ojo}
B.~Abi \textit{et al.} [Muon g-2],
Phys. Rev. Lett. \textbf{126} (2021) no.14, 141801
doi:10.1103/PhysRevLett.126.141801
[arXiv:2104.03281 [hep-ex]].



\bibitem{Wang:2014elb}
L.~Wang and X.~F.~Han,
Phys. Lett. B \textbf{739} (2014), 416-420
doi:10.1016/j.physletb.2014.11.016
[arXiv:1406.3598 [hep-ph]].




\bibitem{Wang:2014sda}
L.~Wang and X.~F.~Han,
JHEP \textbf{05} (2015), 039
doi:10.1007/JHEP05(2015)039
[arXiv:1412.4874 [hep-ph]].


\bibitem{Chaudhuri:2021agl}
A.~Chaudhuri and M.~Y.~Khlopov,
MDPI Physics \textbf{3} (2021) no.2, 275-289
doi:10.3390/physics3020020
[arXiv:2103.03477 [hep-ph]].

 
\bibitem{Chaudhuri:2021rwt}
A.~Chaudhuri, M.~Y.~Khlopov and S.~Porey,
Galaxies \textbf{9} (2021) no.2, 45
doi:10.3390/galaxies9020045
[arXiv:2105.10728 [hep-ph]].


\bibitem{Eberhardt:2020dat}
O.~Eberhardt, A.~P.~Mart\'\i{}nez and A.~Pich,
JHEP \textbf{05} (2021), 005
doi:10.1007/JHEP05(2021)005
[arXiv:2012.09200 [hep-ph]].


\bibitem{Karmakar:2020mds}
S.~Karmakar,
Springer Proc. Phys. \textbf{248} (2020), 193-198
doi:10.1007/978-981-15-6292-1\_23



\bibitem{Dorsch:2016tab}
G.~C.~Dorsch, S.~J.~Huber, K.~Mimasu and J.~M.~No,
Phys. Rev. D \textbf{93} (2016) no.11, 115033
doi:10.1103/PhysRevD.93.115033
[arXiv:1601.04545 [hep-ph]].



\bibitem{Chaudhuri:2017icn}
A.~Chaudhuri and A.~Dolgov,
JCAP \textbf{01} (2018), 032
doi:10.1088/1475-7516/2018/01/032
[arXiv:1711.01801 [hep-ph]].


\bibitem{Bernon:2017jgv}
J.~Bernon, L.~Bian and Y.~Jiang,
JHEP \textbf{05} (2018), 151
doi:10.1007/JHEP05(2018)151
[arXiv:1712.08430 [hep-ph]].



\bibitem{Basler:2018cwe}
P.~Basler and M.~M\"uhlleitner,
Comput. Phys. Commun. \textbf{237} (2019), 62-85
doi:10.1016/j.cpc.2018.11.006


\bibitem{Basler:2020nrq}
P.~Basler, M.~M\"uhlleitner and J.~M\"uller,
Comput. Phys. Commun. \textbf{269} (2021), 108124
doi:10.1016/j.cpc.2021.108124
[arXiv:2007.01725 [hep-ph]].



\bibitem{Basler:2016obg}
P.~Basler, M.~Krause, M.~Muhlleitner, J.~Wittbrodt and A.~Wlotzka,
JHEP \textbf{02} (2017), 121
doi:10.1007/JHEP02(2017)121
[arXiv:1612.04086 [hep-ph]].


\end{thebibliography}
\end{document}